\documentclass[aps,prl,showpacs,amsmath,amssymb,twocolumn,notitlepage]{revtex4-1}
\usepackage{color}
\usepackage{amsmath}
\usepackage{graphicx}
\usepackage{multirow}
\usepackage{float}
\usepackage{bm}

\DeclareMathOperator{\sgn}{sgn} 

\begin{document}

\newcommand{\bn}{{\bm n}}
\newcommand{\bp}{{\bm p}}   
\newcommand{\br}{{\bm r}}
\newcommand{\bk}{{\bm k}}
\newcommand{\bv}{{\bm v}}
\newcommand{\brho}{{\bm{\rho}}}
\newcommand{\bj}{{\bm j}}
\newcommand{\wk}{\omega_{\bf k}}
\newcommand{\nk}{n_{\bf k}}
\newcommand{\eps}{\varepsilon}
\newcommand{\la}{\langle}
\newcommand{\ra}{\rangle}
\newcommand{\be}{\begin{eqnarray}}
\newcommand{\ee}{\end{eqnarray}}
\newcommand{\intl}{\int\limits_{-\infty}^{\infty}}
\newcommand{\dE}{\delta{\cal E}^{ext}}
\newcommand{\SE}{S_{\cal E}^{ext}}
\newcommand{\dsp}{\displaystyle}
\newcommand{\phit}{\varphi_{\tau}}
\newcommand{\p}{\varphi}
\newcommand{\cL}{{\cal L}}
\newcommand{\dphi}{\delta\varphi}
\newcommand{\dbj}{\delta{\bf j}}
\newcommand{\dI}{\delta I}
\newcommand{\dph}{\delta\varphi}
\newcommand{\ua}{\uparrow}
\newcommand{\da}{\downarrow}
\newcommand{\ip}{\{i_{+}\}}
\newcommand{\im}{\{i_{-}\}}
  
\title{Sign reversal of nonlocal response due to electron collisions}

\author{K.E.~Nagaev} 
\affiliation{Kotelnikov Institute of Radioengineering and Electronics, Mokhovaya 11-7, Moscow 125009, Russia}
\affiliation{Institute of Solid State Physics, Russian Academy of
Sciences, 142432 Chernogolovka, Russian Federation}

\date{\today}

\begin{abstract}
Electron-electron collisions are known to cause a nonlocal voltage drop in a presence of current flow. 
The semi-phenomenological theory predicts this drop to be opposite to the direction
of the current in the ballistic regime.
We use a microscopic approach and show that the sign of this drop may be of both 
signs depending on the temperature and the distance between the source and probe contacts. 
The change of sign corresponds to the change of the dominant scattering process from head-on collisions to
backward scattering of electrons. Our results agree with the experimental data.
\end{abstract}

\maketitle

Modern technologies allow a fabrication of conducting microstructures smaller than the elastic mean free path of electrons,
which makes observable the effects of  electron--electron scattering in them. Though this scattering does not contribute 
to the resistivity of a homogeneous conductor with a 
parabolic spectrum, it affects the transport properties of finite-size systems in  different 
ways. These effects were investigated in a big number of papers in the hydrodynamic regime where the mean free path of 
electron--electron scattering $l_{ee}$ is much smaller than the characteristic size of the structure. Among others, they 
included a temperature-dependent resistance of conducting channels with rough boundaries known as the Gurzhi effect 
\cite{Gurzhi68,deJong95,Gusev18}, but also  a negative nonlocal resistance \cite{Levitov16,Bandurin16,Braem18}, which 
represents a current-induced voltage drop in the direction opposite to the current flow. This negative resistance was even 
deemed a signature of the hydrodynamic regime \cite{Bandurin16}.

Along with this, the effects of two-particle collisions were also investigated in the ballistic regime where the 
characteristic size of the structure is much smaller than $l_{ee}$. In wide ballistic contacts, these collisions 
were shown to result in a contribution to the current that linearly grows with temperature. This contribution was 
theoretically predicted in Refs. \cite{Nagaev08,Nagaev10} and experimentally observed in Refs. \cite{Renard08,Melnikov12}. 
More recently, the negative nonlocal resistance in the ballistic regime was obtained in Ref. \cite{Shytov18} 
using the Boltzmann equation in the semi-phenomenological approximation of a single electron--electron scattering time.  
However it was observed \cite{Braem18,Bandurin18} that at lower temperatures this resistance becomes positive. 
These authors explained the change of sign by  the finite-size effect and attributed it to reflections of the electrons from the 
opposite boundary of the conductor.

This paper presents a microscopic calculation of the nonlocal response of a ballistic conductor with two-particle 
scattering. In this approach, the change of sign of the response takes place at low temperatures even in the absence 
of boundary reflections as a result of competition between different scattering processes in the electron--electron 
collision integral. The temperature dependence of the effect also appears to be different from that obtained in the 
semi-phenomenological approximation. 

\begin{figure}
 \includegraphics[width=1.0\columnwidth]{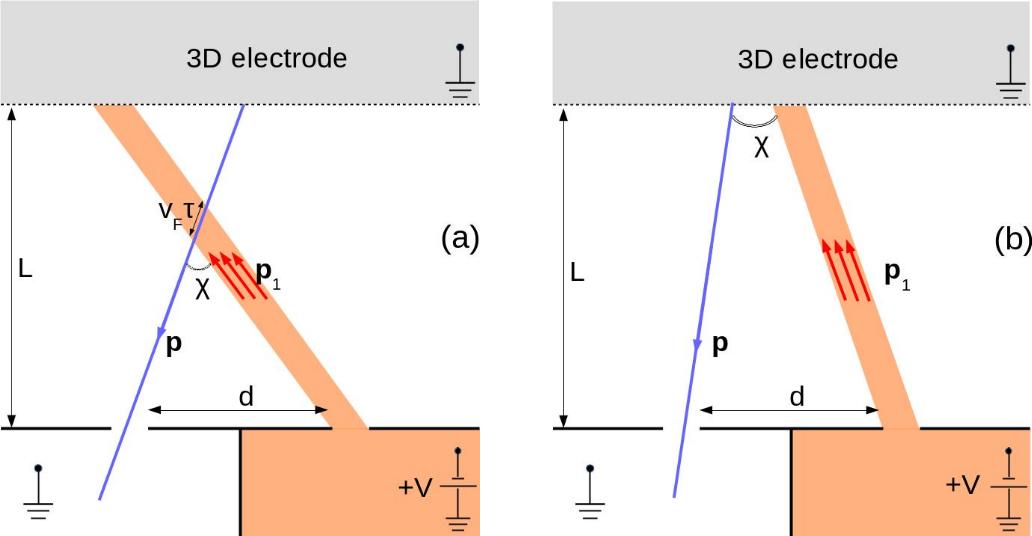}
 \caption{\label{fig:system} The layout of the system. (a) Nonequilibrium electrons from the source contact with momentum
 $\bp_1$ cross the trajectory of the electron incident on the probe contact with momentum $\bp$ if the angle $\chi$ between $-\bp$ 
 and $\bp_1$ is not too small. (b) If $\chi$ is too small, the injected beam does not cross the trajectory of incident electron.
 }
 \vspace*{-5mm}
\end{figure}

The sketch of the system considered is shown in Fig.~\ref{fig:system}. It represents a conducting plane separated by thin 
insulating barriers into three parts, i.~e. one grounded half-plane, one grounded quarter-plane, and one more 
quarter-plane kept at a constant voltage $V$. The half-plane is connected with the grounded quarter-plane by
the probe contact of width $a\gg\lambda_F$ and with the voltage-biased quarter-plane, by the source contact of width 
$b\gg\lambda_F$. The distance between the contacts is $d \gg {\rm max}(a,b)$. A perfect 3D electrode is attached to the
conducting half-plane at a distance $L\gg d$ from the barrier.
It is assumed that the electron-electron scattering is so weak that $l_{ee}\gg L$.  If $V=0$, the fluxes of 
equilibrium electrons from both
sides of the probe contact compensate each other and the net electric current through it is zero. At nonzero bias,
nonequilibrium electrons are injected into the grounded half-space through the source contact and collide with equilibrium 
electrons incident on the probe contact so that their flux is changed, which results in a net electric current of either
positive or negative sign. If the circuit is disconnected and the current flow through the probe contact is forbidden,
this results in the nonlocal voltage that compensates it.

To calculate the current through the probe contact, we use an approach similar to that of Ref. \cite{Nagaev08}. The 
distribution function of electrons $f(\bp,\br)$ in all the three parts of the plane obey the Boltzmann equation
\cite{*[{The large number of conducting channels in the contacts allows us to neglect the corrections to the conductance
from quantum interference, which are of the order $e^2/\hbar$, see }][{}] Beenakker97}
\be
 \frac{\partial f}{\partial t}
 +
 \bv\,\frac{\partial f}{\partial\br}
 +
 e{\bm E}\,\frac{\partial f}{\partial\bp}
 =
 {I}_{ee},
 \label{Boltz}
\ee
where ${\bf E} = -\nabla\varphi$ is the electric field. The electron--electron
collision integral is given by
\begin{multline}
 \hat{I}_{ee}(\bp)
 =
 \frac{\alpha_{ee}}{\hbar\nu_0^{2}}
 \int\frac{d^2p_1}{(2\pi\hbar)^2}
 \int\frac{d^2p_2}{(2\pi\hbar)^2}
 \int d^2p_3
\\ \times
 \delta(\bp + \bp_1 - \bp_2 - \bp_3)\,
 \delta( \eps_{\bp} + \eps_{\bp_1} - \eps_{\bp_2} - \eps_{\bp_3} )
\\ \times
 \Bigl\{
  [1 - f(\bp)]\,[1 - f(\bp_1)]\, f(\bp_2)\, f(\bp_3)\,
\\  -
  f(\bp)\, f(\bp_1)\, [1 - f(\bp_2)]\, [1 - f(\bp_3)]
 \Bigr\},
 \label{Iee}
\end{multline}
where $\alpha_{ee}$ is the dimensionless parameter of electron--electron scattering and $\nu_0 = m/\pi\hbar^2$ is the
2D density of states. 

In the absence of electron-electron scattering, the nonequilibrium electrons injected through the source contact would travel 
along their classical trajectories  to the depth of the grounded half-plane, and no current would flow through the probe
contact. Therefore the nonlocal current flowing into the   grounded half-plane may be written in the form
\be
 I_n = e\int_a d\rho \int\frac{d^2p}{(2\pi\hbar)^2}\,v_{\perp}\,\delta f(\bp,\brho),
 \label{I-1}
\ee
where $\brho$ labels points within the probe contact and $\delta f$ is the correction to the distribution of 
noninteracting electrons $f^{(0)}(\bp,\br)$ from two-particle
scattering. It may be obtained by integrating the collision integral in Eq. \eqref{Boltz} along the trajectories of electrons 
that come to point $\brho$ with momentum $\bp$ from the 3D electrode. Because of the condition $E_F \gg {\rm max}(eV,T)$ one 
may neglect the change of electron velocity caused by the electric field, so \cite{Nagaev08}
\be
 \delta f(\bp,\brho)=
 \int_0^{\infty} d\tau\,
  \hat{I}_{ee}(\bp,\brho-\tau\bv),
 \label{df}
\ee
where $\tau$ is the time of travel to point $\brho$. 

\begin{figure}
\includegraphics[width=0.8\columnwidth]{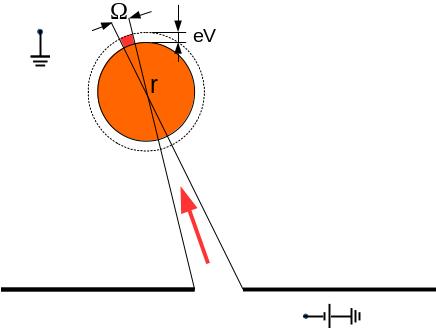}
\caption{\label{fig:distribution} The distribution function of non-interacting electrons at a given point $\br$ in the 
grounded half-plane. The electrons injected through the source contact form a bump on the Fermi surface.
}
\vspace*{-5mm}
\end{figure}

\begin{figure}
\includegraphics[width=0.9\columnwidth]{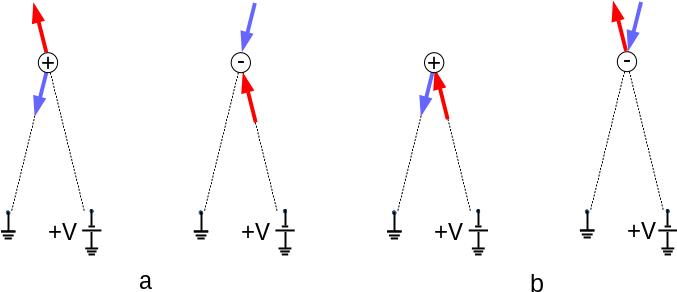}
\caption{\label{fig:processes} The scattering processes that contribute to the collision integral. (a) Head-on 
collisions and (b) backward scattering.  Red arrows denote the injected nonequilibrium electrons, and blue arrows denote
electrons incident on the probe contact. The "plus" and "minus" signs denote the sign of the corresponding contribution
to the nonlocal current.}
\vspace*{-5mm}
\end{figure}

To the first approximation in $\alpha_{ee}$, the inelastic correction Eq. \eqref{df} may be calculated using the
distribution function of noninteracting electrons. As the total energy of electron is conserved during its motion 
along the classical trajectory, $f^{(0)}(\bp,\br)$ depends only on the electrode from which the electron trajectory 
originates. Denote the angular domain  that contains all the momenta of electrons that came to point $\br$ of the 
grounded half-plane through the source contact by $\Omega(\br)$ \cite{Kulik77}, see Fig.~\ref{fig:distribution}. With 
this notation,
\be
 f^{(0)}(\bp,\br) =
 \begin{cases}
  f_0(\eps_{\bp}+e\p(\br)-eV), & \bp\in\Omega(\br)
  \\
  f_0(\eps_{\bp}+e\p(\br)),    & \bp\notin\Omega(\br)
 \end{cases},
 \label{f0}
\ee
where $f_0$ is the Fermi distribution. The collision integral Eq. \eqref{Iee} is nonzero only if at least one of the
momenta lies in $\Omega(\br)$. On the other hand, the integral \eqref{df} is dominated by large $\tau$ and hence
small $\Omega$, so the contributions from scattering processes involving more than one momentum in $\Omega(\br)$ may be 
neglected. In view of Eqs. \eqref{df} and \eqref{Iee}, the nonlocal current Eq. \eqref{I-1} may be written in the form
\begin{multline}
 I_n = ea\,\frac{\alpha_{ee}}{\hbar\nu_0^2} \iiiint d\eps\,  d\eps_1\, d\eps_2\, d\eps_3\,
 \iint \frac{d^2p}{(2\pi\hbar)^2}\,\frac{d^2p_1}{(2\pi\hbar)^2}
  \\ \times
 \delta(\eps_{\bp}-\eps)\,\Theta(v_{\perp})\,v_{\perp}
 \delta(\eps_{\bp_1}-\eps_1)\,\tau_m(\bp,\bp_1)
 \\ \times
 \bigl[
  \delta(\eps+\eps_1-\eps_2-\eps_3)\,A(\eps_2,\eps_3,\bp+\bp_1)\,F(\eps,\eps_1;\eps_2,\eps_3)
 \\ 
  -\delta(\eps-\eps_1+\eps_2-\eps_3)\,A(\eps_2,\eps_3,\bp-\bp_1)\,F(\eps_3,\eps_1;\eps_2,\eps) 
 \\
  -\delta(\eps-\eps_1+\eps_3-\eps_2)\,A(\eps_3,\eps_2,\bp-\bp_1)\,F(\eps_2,\eps_1;\eps_3,\eps)\bigr],
 \label{I-2}
\end{multline}
where
\be
 \tau_m(\bp,\bp_1) = \int_0^{\infty} d\tau\,\Theta(\bp_1 \in {\bm\rho} - \tau\,{\bm v}),
 \label{tau-1}
\ee
is the effective dwell time of electrons incident on the probe contact with momentum $\bp$ in a beam of nonequilibrium 
electrons with momentum $\bp_1$ injected through the source contact, 
\begin{multline}
 A(\eps_2,\eps_3,{\bm Q})= \frac{1}{(2\pi\hbar)^2} \int d^2p' \int d^2 p''\,\delta(\bp' \pm \bp'' - {\bm Q})
 \\ \times
 \delta(\eps_{\bp'}-\eps_2)\,\delta(\eps_{\bp''}-\eps_3)
 \label{A-1}
\end{multline}
is the phase volume made available for the corresponding scattering process by the momentum conservation, and
\begin{multline}
 F(\eps,\eps_1;\eps_2,\eps_3) = 
 [1-f_0(\eps)][1-f_0(\eps_1-eV)]\,f_0(\eps_2)\,f_0(\eps_3)
\\ -
 f_0(\eps)\,f_0(\eps_1-eV)\,[1-f_0(\eps_2)][1-f_0(\eps_3)]
 \label{F}
\end{multline}
is the distribution-dependent factor. The first term in square brackets in Eq. \eqref{I-2} describes the scattering of 
an electron incident on the probe contact by a nonequilibrium electron
or reverse process (Fig.~\ref{fig:processes}a). The second  and the third terms are equivalent and correspond to the scattering 
of a nonequilibrium electron into the probe contact by  an equilibrium electron or reverse process (Fig.~\ref{fig:processes}b). 
At long distances  from the contacts, these terms correspond to head-on collisions and backward scattering, respectively. They 
are of opposite signs and the competition between them determines the overall sign of nonlocal current.

\begin{figure}
 \includegraphics[width=0.8\columnwidth]{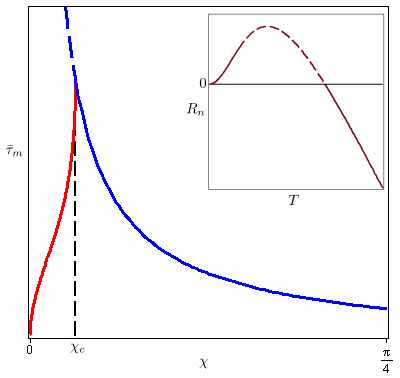}
 \caption{\label{fig:dwell} The $\bar\tau_m(\chi)$ curve for $L/d=10$. The blue dashed curve shows this dependence
  in the limit $L\to\infty$. The suppression takes place at $\chi <\chi_c \sim d/L$. The inset shows the approximate 
  theoretical $R_n(T)$ dependence.}
  \vspace*{-5mm}
\end{figure}

It is convenient to describe the dwell time $\tau_m$ in terms of the angle $\phi$ between the normal to the
insulating barrier and the trajectory of incident electron and the angle $\chi$ between $-\bp$ and $\bp_1$.
If the electron--electron scattering took place in the whole grounded half-plane, it would increase infinitely at
$\chi\to 0$, see Fig.~\ref{fig:system}a. But if the trajectories of the incident and the injected electrons intersect
further from the barrier than $L$, the scattering does not take place as shown in Fig.~\ref{fig:system}b . Therefore the 
$\phi$-weighted dwell time sharply falls down to zero at $\chi<d/L$ and may be approximately written in the form
\begin{multline}
 \bar\tau_m(\chi) = \int_{-\pi/2}^{\pi/2} \frac{d\phi}{2\pi}\,\cos\phi\,\tau_m(\phi,\chi)
 = \frac{1}{4\pi}\,\frac{b}{v_F}\,\Theta(\chi-d/L)
 \\ \times
 \bigl[(\pi-\chi)\,\cot\chi + 1].
 \label{tau-2}
\end{multline}
A sketch of exact $\bar\tau_m(\chi)$ is shown in Fig.~\ref{fig:dwell}.
The phase-space factors $A$ in Eq. \eqref{I-2} also have singularities at $\bp+\bp_1=0$ of the form 
\be
 A(\eps_2, \eps_3, \bp\pm\bp_1) = \frac{(2\pi\hbar v_F)^{-2}\,\Theta(\eta_{\pm})}{\cos(\chi/2)\,\sqrt{\eta_{\pm}}},
 \label{A-2}
\ee
where $\eta_{+} = \sin^2(\chi/2)+(\eps_2-\eps)(\eps_1-\eps_2)/4E_F^2$ and $\eta_{-} = \sin^2(\chi/2) + (\eps_1-\eps_2)/E_F$
account for the thermal smearing at $\chi=0$.
The singularity in the first
term  of Eq. \eqref{I-2} is well known and results in an additional logarithmic factor in the rate of
electron--electron scattering \cite{Chaplik71,Hodges71,Giuliani82}. The singularities in the two last terms are smaller at 
$T\ll E_F$ and normally less important.

The relative magnitude of the terms in Eq. \eqref{I-2} depends on the ratio between $d/L$ and $T/E_F$. 
If the temperature is high or the contacts are so close that $d/L\ll T/E_F$, both the cutoff in the dwell time 
and the thermal smearing of the phase volume are essential, and therefore upon the integration over $\chi$, the most singular 
contribution at low temperatures from the first term 
\begin{multline}
 I_n = -e\,\frac{\alpha_{ee}\nu_0 }{32\pi^2\hbar}\,ab\,\ln\!\left(\frac{d}{L}\right)
 \iiiint d\eps\,d\eps_1\,d\eps_2\,d\eps_3
 \\ \times
 \delta(\eps+\eps_1-\eps_2-\eps_3)\,
 \frac{\Theta\bigl[(\eps_2-\eps)(\eps_1-\eps_2)\bigr]}{\sqrt{(\eps_2-\eps)(\eps_1-\eps_2)}}
 \\ \times
 F(\eps,\eps_1;\eps_2,\eps_3)
 \label{I-H}
\end{multline}
is $(E_F/T)^{1/2}$ times larger than that from the second and third terms. In the limit $eV \ll T$, Eq. \eqref{I-H}
reduces to
\be
 I_n = \frac{C_0 \alpha_{ee}}{32\pi^2}\,\frac{e^2}{\hbar}\,V ab\,\nu_0 T
 \,\ln\!\left(\frac{L}{d}\right),
 \quad C_0 \approx 3.46.
 \label{I-HT}
\ee
This expression is very similar to the one obtained in Ref.~\cite{Nagaev08} for the inelastic correction to the current
through a ballistic contact except that the square of the contact width is replaced by the product of widths of source and 
probe contacts. At the same time, the nonlocal current Eq. \eqref{I-HT} exhibits a linear temperature dependence, while in
Ref. \cite{Shytov18} it is quadratic with a similar logarithmic factor. One power of $T$ is eliminated due to the superposition 
of the two singularities in the dwell time and in the  phase volume available for the scattering of electrons with almost 
opposite momenta. In the opposite limit $eV \gg T$, $T$ is effectively replaced by $eV$, so
\be
 I_n = \sgn V\,\frac{(4-\pi)\, \alpha_{ee}}{64\pi^2}\,\frac{e^2}{\hbar}\,V^2  ab\,\nu_0
 \,\ln\!\left(\frac{L}{d}\right).
 \label{I-HV}
\ee

If 
$d/L\gg {\rm max}(T,eV)/E_F$, finite dwell time cuts off the integration at small $\chi$ before the thermal
smearing of the Fermi surface comes into play, so the phase-volume factors may be written simply as
$A = (2\pi^2\hbar^2 v_F^2\sin\chi)^{-1}$. Therefore all the three terms in Eq. \eqref{I-2} give the contributions to $I_n$
with the same absolute values, but the signs of the two last terms are opposite to the first one, so the total nonlocal current 
\be
 I_n = -\frac{\alpha_{ee}}{192\pi^2}\,\frac{e^2}{\hbar}\,V\,ab\nu_0\, \frac{e^2V^2+4\pi^2T^2}{E_F}\,\frac{L}{d}
\label{dI-L}
\ee
changes the sign. Therefore at $eV \ll T$, the nonlocal current exhibits a nonmonotonic temperature dependence. 
In the absence of boundary 
reflections, it starts with a quadratic growth at zero temperature, then reaches a maximum and decreases to become
negative, while its absolute value linearly grows with temperature until $l_{ee}$ becomes smaller than $L$.

If the quarter-plane on the opposite side of the probe contact is electrically isolated instead of being grounded, a compensating
nonlocal voltage of opposite sign to the calculated nonlocal current arises. The quantity measured in experiments of Braem et al.
\cite{Braem18} was $R_n=V_n/I$, where $V_n=-R_p I_n$ is the voltage drop across the probe contact with resistance $R_p$
and $I=V/R_s$ is the current through the source contact with resistance $R_s$. 
The geometry investigated there is different from ours and the nonlocal response 
$R_n$ contains a significant contribution from electron scattering at the opposite boundary of the channel. However this 
contribution should be temperature-independent as long as the width of the channel is smaller than $l_{ee}$. If one subtracts from 
$R_n(T)$ its value at $T=0$, the resulting curves are in a good agreement with our predictions shown in Fig~\ref{fig:dwell},
inset. In particular, these curves exhibit clear 
maxima at low temperatures. Note also that the negative-slope linear portion of $R_n(T)$ corresponds to the temperature range about 
$T= 2$ K where a positive correction to the conductance of a contact from electron--electron scattering was observed 
\cite{Renard08,Melnikov12}.
At these temperatures,  $l_{ee}\sim 10\,\mu$m while the width of the channel in Ref. \cite{Braem18} is 5 $\mu$m, so the 
assumption of ballistic regime is justified.

In the lowest approximation in $\alpha_{ee}$, one may calculate the resistances $R_p$ and $R_s$ using the Sharvin 
expressions $\pi^2\hbar^2/(e^2 p_F a)$ and $\pi^2\hbar^2/(e^2 p_F b)$ \cite{Sharvin65}. Assuming that
the electron concentration is $n = 1.2\times 10^{11}$ cm$^{-2}$ and mobility is $\mu = 6\times 10^6$~cm$^2$/(V$\cdot$s) as in
Ref. \cite{Braem18}, and that $\alpha_{ee} \sim 1$ \cite{Altshuler85}, one obtains from Eq. \eqref{I-HT} the estimate of 
temperature slope of $R_n$ normalized to the sheet resistance $\rho$ of the conducting plane about -2 K$^{-1}$. This is of the same 
order of magnitude as the temperature slope of $R_n/\rho$ in Ref. \cite{Braem18}.
It would be of interest to set up an experiment in which the reflections from the boundaries would play no role, so that there 
would be no need to subtract their contribution from the nonlocal response, and to investigate the low-temperature portions
of $R_n(T)$ in more detail. 
These measurements could provide an insight into microscopic processes of 
electron--electron scattering in 2D conductors

In summary, the spatial confinement changes the relative importance of different scattering channels in the
electron--electron collision integral. While for large systems the singularities in the available phase-space volume are smeared
by the finite temperature and head-on collisions dominate over backward scattering, for small enough systems they are
cut off by a finite dwell time of an electron in the system and the backward scattering becomes the dominating process.
This results in the change of sign of the nonlocal response, which may be experimentally observable.

I am grateful to Vadim Khrapai for a very motivating discussion.
This work was supported by Russian Science Foundation (Grant No 19-12-00326).

\bibliography{ee-refs}

\end{document}